%Paper: hep-ph/9503495
%From: Fumihiko TOYODA <ftoyoda@fuk.kindai.ac.jp>
%Date: Mon, 3 Apr 1995 12:07:00 +0900

% Chiral Charge Flux and Electroweak Baryogenesis
%
%            -----        CORRECTED VERSION      -----
%
%       required macros : phyzzx
% ----------------------------------------------------------------
\input phyzzx
%%%%%%%%%%%%%
%%--- local extension
%%
\def\gtsim{\mathrel{\hbox{\raise0.2ex
\hbox{$>$}\kern-0.75em\raise-0.9ex\hbox{$\sim$}}}}
\def\ltsim{\mathrel{\hbox{\raise0.2ex
\hbox{$<$}\kern-0.75em\raise-0.9ex\hbox{$\sim$}}}}
%                               %  inequality with sim-simbol
\catcode`\@=11
%\newfam\mibfam
% Extra fonts
%
\font\sevensl =cmsl8 at 7pt
\font\sevenit =cmti7
%
%\font\seventeenmib =cmmib10 scaled\magstep3 \skewchar\seventeenmib='177
%\font\fourteenmib =cmmib10 scaled\magstep2 \skewchar\fourteenmib='177
%\font\twelvemib =cmmib10 scaled\magstep1 \skewchar\twelvemib='177
%\font\tenmib =cmmib10 \skewchar\tenmib='177
%\font\ninemib =cmmib9 \skewchar\ninemib='177
%\font\sixmib =cmmib6 \skewchar\sixmib='177
%
% Redefinition of fourteenf@nts
\def\fourteenf@nts{\relax
\textfont0=\fourteenrm \scriptfont0=\tenrm
\scriptscriptfont0=\sevenrm
\textfont1=\fourteeni \scriptfont1=\teni
\scriptscriptfont1=\seveni
\textfont2=\fourteensy \scriptfont2=\tensy
\scriptscriptfont2=\sevensy
\textfont3=\fourteenex \scriptfont3=\twelveex
\scriptscriptfont3=\tenex
\textfont\itfam=\fourteenit \scriptfont\itfam=\tenit
\textfont\slfam=\fourteensl \scriptfont\slfam=\tensl
\textfont\bffam=\fourteenbf \scriptfont\bffam=\tenbf
\scriptscriptfont\bffam=\sevenbf
%\textfont\mibfam=\fourteenmib \scriptfont\mibfam=\tenmib
\textfont\ttfam=\fourteentt
\textfont\cpfam=\fourteencp }
% Redefinition of twelvef@nts
\def\twelvef@nts{\relax
\textfont0=\twelverm \scriptfont0=\ninerm
\scriptscriptfont0=\sixrm
\textfont1=\twelvei \scriptfont1=\ninei
\scriptscriptfont1=\sixi
\textfont2=\twelvesy \scriptfont2=\ninesy
\scriptscriptfont2=\sixsy
\textfont3=\twelveex \scriptfont3=\tenex
\scriptscriptfont3=\tenex
\textfont\itfam=\twelveit \scriptfont\itfam=\nineit
\textfont\slfam=\twelvesl \scriptfont\slfam=\ninesl
\textfont\bffam=\twelvebf \scriptfont\bffam=\ninebf
\scriptscriptfont\bffam=\sixbf
%\textfont\mibfam=\twelvemib \scriptfont\mibfam=\ninemib
%\scriptscriptfont\mibfam=\sixmib
\textfont\ttfam=\twelvett
\textfont\cpfam=\twelvecp }
% Redefintion of tenf@nts
\def\tenf@nts{\relax
\textfont0=\tenrm \scriptfont0=\sevenrm
\scriptscriptfont0=\fiverm
\textfont1=\teni \scriptfont1=\seveni
\scriptscriptfont1=\fivei
\textfont2=\tensy \scriptfont2=\sevensy
\scriptscriptfont2=\fivesy
\textfont3=\tenex \scriptfont3=\tenex
\scriptscriptfont3=\tenex
\textfont\itfam=\tenit \scriptfont\itfam=\sevenit
\textfont\slfam=\tensl \scriptfont\slfam=\sevensl
\textfont\bffam=\tenbf \scriptfont\bffam=\sevenbf
\scriptscriptfont\bffam=\fivebf
%\textfont\mibfam=\tenmib
\textfont\ttfam=\tentt
\textfont\cpfam=\tencp }
\Twelvepoint
\catcode`\@=12
%%
%%--- end of local extension
%%
\tolerance=9999
\overfullrule=0pt

\def\del{\partial}
\def\dslash{\del\kern-0.55em\raise 0.14ex\hbox{/}}

%\def\bfx{{\mib x}}
%\def\bfpT{{{\mib p}_T}}

%
% Page style definition
%
\def\papersize{\hsize=37pc \vsize=52pc \hoffset=1.5mm \voffset=1pc
\advance\hoffset by\HOFFSET \advance\voffset by\VOFFSET
\pagebottomfiller=0pc
\skip\footins=\bigskipamount \normalspace }
\papersize
\date={April 3,  1995}
\Pubnum={}
\Pubtype={}
\titlepage
\centerline{}
\vskip -1.2cm
\title{\fourteenbf Chiral Charge Flux and Electroweak Baryogenesis}
\bigskip\bigskip
\centerline{Koichi Funakubo${}^{1)}$\footnote{\#a}
{e-mail: \ funakubo@cc.saga-u.ac.jp},\ \ \  Akira Kakuto${}^{2)}$%
\footnote{\#b}{e-mail: \ kakuto@fuk.kindai.ac.jp},
\ \ Shoichiro Otsuki${}^{2)}$\footnote{\#c}{e-mail: \
otks1scp@mbox.nc.kyushu-u.ac.jp},}
\centerline{Kazunori Takenaga${}^{3)}$\footnote{\#d}%
{e-mail: \ take1scp@mbox.nc.kyushu-u.ac.jp}
\ \ and\ \ \  Fumihiko Toyoda${}^{2)}$\footnote{\#e}%
{e-mail: \ ftoyoda@fuk.kindai.ac.jp}}
\bigskip
\centerline{${}^{1)}$\it Department of Physics, Saga University,
Saga 840 Japan}
\centerline{${}^{2)}$\it Department of Liberal Arts, Kinki University in
Kyushu, Iizuka 820 Japan}
\centerline{${}^{3)}$\it Yukawa Institute for Theoretical Physics, Kyoto
University,
Kyoto 606-01 Japan}
\bigskip\bigskip
%%%%%%%%%
\abstract{
By treating CP-violating interaction of the electroweak bubble wall as a
perturbative term, chiral charge flux through
the bubble wall is estimated.
It is found that the absolute value of the flux $F_Q$ has
a sharp peak at $m_0 \sim a \sim T$
with $F_Q/(u T^3) \sim 10^{-3}\, (Q_L-Q_R)\,\Delta \theta$.
Here $m_0$ is the fermion mass, $1/a$ is the wall thickness, $T$ is the
temperature at which the bubbles are growing, $u$ is the wall velocity,
$Q_{L(R)}$ is the chiral charge of the relevant left(right)-handed
fermion and $\Delta\theta$ is the measure of CP violation.

\vfill\eject
\chapter{Introduction}
 Provided that the electroweak phase transition is first order with the
 bubble formation${}^{1)2)}$,
 a fermion gets a complex mass
 through the transition in the presence of CP violation:
 $m(z) = m_R(z) + im_I(z)
 \equiv m_0(f(z)+ig(z)) $, where $z$ is the coordinate perpendicular
 to the wall, $f(z)$ gives the
 profile of the wall, $m_0$ is the fermion mass at the temperature
 in question, and the imaginary term $g(z)$ yields CP violation
 necessary for baryogenesis${}^{3)}$.
\par
In a previous paper, we have examined fermion scattering off the
CP-violating bubble wall by treating $g(z)$ as a perturbative term,
and have derived the CP-violating effect
$\Delta R \equiv R^s_{R\rightarrow L} - \bar R^s_{R\rightarrow L}
\propto \Delta\theta $ in the kink-type background of
$f(z)=(1+\tanh(az))/2$
 (distorted wave Born approximation---DWBA)${}^{4)}$\footnote{\dag}
 {Reference 4) is referred to as I hereafter.}.
 Here $R^s_{R\rightarrow L}$($\bar R^s_{R\rightarrow L}$)
is the reflection coefficient for the right-handed fermion
(anti-fermion) incident from the symmetric phase region and reflected
as the left-handed one, and $\Delta\theta$ is the measure of CP-violation.
$\mid\Delta R\mid$ is found to decrease exponentially as
$a/m_0$ decreases, where $1/a$ is the thickness of the
bubble wall.\footnote{\dag\dag}{Because of some errors in our numerical
calculations,
figures and table in I are incorrect.}
This means that, given $1/a$, the heavier fermion whose Compton wave length
is less than the wall thickness gives the smaller contribution to
$\mid\Delta R\mid$. In other words, a large Yukawa
coupling does not necessarily cause a large effect of CP violation in
electroweak baryogenesis.\par
In this paper, we analyze the chiral charge flux $F_Q \propto \Delta\theta $
just in front of the bubble wall in the symmetric phase region.
$F_Q$ is given by integrating $\Delta R$ weighted by the fermion
flux densities in Sec.2.
In Sec.3 numerical estimates of $F_Q$ are given. $F_Q$ has a sharp peak at
$m_0\sim a\sim T$,
 where $T$ is the temperature
at which the bubbles are growing.
At the peak we find $F_Q/(u T^3) \sim 10^{-3}\,(Q_L-Q_R)\,\Delta\theta$,
where $u$ is the wall velocity and $Q_{L(R)}$ is the chiral charge of
 the relevant fermion.
 Several remarks on electroweak baryogenesis are given in Sec.4.
% end of Introduction
%
%
\chapter{Chiral Charge Flux through Bubble Wall}
Suppose that a left(right)-handed fermion of the species $i$ with chiral
 charge $Q^i_{L(R)}$, which is conserved in the symmetric phase,
 is incident to the wall from the symmetric phase region.
The change of the expectation value of the chiral charge
in the symmetric phase region, which has been brought by
the reflection and transmission of the fermions, is
given by
$$
\eqalign{
\Delta Q^s_i
=&\bigl[ (Q^i_R-Q^i_L)R^s_{L\rightarrow R}+(-Q^i_L+Q^i_R)\bar R^s_{R\rightarrow
L} \cr
 &\quad
     +(-Q^i_L)(T^s_{L\rightarrow L}+T^s_{L\rightarrow R})
     -(-Q^i_R)( \bar T^s_{R\rightarrow L}+\bar T^s_{R\rightarrow
R})\bigr]f^s_{Li} \cr
+&\bigl[(Q^i_L-Q^i_R)R^s_{R\rightarrow L}+(-Q^i_R+Q^i_L)\bar R^s_{L\rightarrow
R} \cr
 &\quad
     +(-Q^i_R)(T^s_{R\rightarrow L}+R^s_{R\rightarrow R})
     -(-Q^i_L)(\bar T^s_{L\rightarrow L}+\bar T^s_{L\rightarrow
R})\bigr]f^s_{Ri}. \cr}
\eqno(2.1)
$$

%$$  \equiv (Q^i_L-Q^i_R)[(R^s_{R\rightarrow L}-R^s_{L\rightarrow R})
%f^s_i+T^{(0)}\Delta f^s_i].\eqno(2.1)$$
\noindent The notations would be obvious: $T$'s are the transmission
coefficients
and $f^s_{L(R)}$ is the left(right)-handed fermion flux density
in the symmetric phase region, and $\bar f_L^s=f_R^s$ has been
taken into account.
Making use of unitarity relations examined in I, we obtain
$$\Delta Q^s_i = (Q^i_L-Q^i_R)[(R^s_{R\rightarrow L}-R^s_{L\rightarrow R})
f^s_i+T^{(0)}(f^s_{Li}-f^s_{Ri})],\eqno(2.2)$$
where $f^s_i \equiv (1/2)(f^s_{Li}+f^s_{Ri})$ and $T^{(0)}$ is the
transmission coefficient in the absence of CP violation.
Note that
$$\Delta R \equiv R^s_{R\rightarrow L} -
\bar R^s_{R\rightarrow L} \eqno(2.3)$$
and
$ \Delta f^s_i \equiv f^s_{Li}-f^s_{Ri} $
are quantities of $O(\Delta\theta)$.
In a similar way to (2.1), the change of the expectation value of
the chiral charge brought
by the transmission of the fermions incident from
the broken phase region is
$$
\eqalign{
\Delta Q^b_i
&=Q^i_L(T^b_{L\rightarrow L}f^b_{Li}+T^b_{R\rightarrow L}f^b_{Ri})
+Q^i_R(T^b_{L\rightarrow R}f^b_{Li}+T^b_{R\rightarrow R}f^b_{Ri}) \cr
&+(-Q^i_L)( \bar T^b_{R\rightarrow L}f^b_{Li}+
\bar T^b_{L\rightarrow L}f^b_{Ri})
+(-Q^i_R)( \bar T^b_{R\rightarrow R}f^b_{Li}+
\bar T^b_{L\rightarrow R}f^b_{Ri}).\cr}
\eqno(2.4)
$$
In addition to unitarity relations in the broken phase region,
$$
\eqalign{
T^b_{L\rightarrow L}&+T^b_{L\rightarrow R}+R^b_{L\rightarrow L}+
R^b_{L\rightarrow R}=1,\cr
T^b_{R\rightarrow L}&+T^b_{R\rightarrow R}+R^b_{R\rightarrow L}+
R^b_{R\rightarrow R}=1,\cr}
\eqno(2.5)
$$
the following reciprocity relations hold${}^{5)}$:
$$
\eqalign{
T^b_{L\rightarrow L}&+T^b_{R\rightarrow L}=T^s_{R\rightarrow L}+
T^s_{R\rightarrow R}(=1-R^s_{R\rightarrow L}),\cr
T^b_{L\rightarrow R}&+T^b_{R\rightarrow R}=T^s_{L\rightarrow L}+
T^s_{L\rightarrow R}(=1-R^s_{L\rightarrow R}).\cr
}
\eqno(2.6)
$$
Thanks to these relations, we obtain
$$
\Delta Q^b_i=-(Q^i_L-Q^i_R)\bigl[
(R^s_{R\rightarrow L}-R^s_{L\rightarrow R})
f^b_i-
{{\sqrt{p_L^2-m_0^2}}\over {p_L}}T^{(0)}
(f^b_{Li}-f^b_{Ri})
\bigr],\eqno(2.7)$$
where $p_L$ is the longitudinal momentum of the fermions to the bubble wall.
Putting all these together, the chiral charge flux just in front of the wall
in the symmetric phase region is given by
$$
F^i_Q={{Q^i_L-Q^i_R}\over{4\pi^2 \gamma}}\int_{m_0}^{\infty}
dp_L\int_0^{\infty} p_Tdp_T
\bigl[f^s_i(p_L,p_T)-f^b_i(p_L,p_T)\bigr]\Delta R(a/m_0, p_L).
\eqno(2.8)
$$
%$f^{s(b)}_i$ is a free fermion distribution function in the
%symmetric(broken) phase and is given by
Here
$$
\eqalign{
f^s_i&=(p_L/E)(\exp[\gamma(E-up_L)/T]+1)^{-1},\cr
f^b_i&=(p_L/E)(\exp[\gamma(E-u\sqrt{p^2_L-m^2_0})/T]+1)^{-1},\cr}
\eqno(2.9)
$$
are the fermion flux densities in the symmetric and broken phase
regions respectively,
 $u$ is the wall velocity, $\gamma=1/\sqrt{1-u^2}$,
$E=\sqrt{p^2_L+p^2_T}$, $p_T$ being the transverse momentum
of the fermions along the wall. $\Delta R$ is defined in (2.3).
In (2.8) we have ignored the effect of CP violation on the
fermion flux densities (the second terms in (2.2) and (2.7).
See the discussions in Sec.4).
%
%end of section 2
%
\chapter{Numerical Estimates of Chiral Charge Flux}
Let the fermion mass be
$$ m(az)\equiv m(x)= m_0(f(x)+ig(x))\qquad {\rm with}\qquad x \equiv az.
\eqno(3.1)$$
We take the kink-type profile of the wall with the thickness $1/a$:
$$ f(x)=(1+\tanh x)/2, \eqno(3.2)$$
and apply DWBA to the imaginary part $g(x) \propto \Delta \theta$ to
calculate the CP-violating effect $\Delta R \propto \Delta \theta$.
The next order corrections would be $O((\Delta \theta)^3)$
since $\Delta R$ is a CP-odd quantity.

The profile of the complex mass $m(x)$ should be determined by
solving the equations of motion of the gauge-Higgs system with
the effective potential at the temperature considered.
In the case of moderate first order phase transition,
we expect the modulus of the mass to be the kink-type profile. Its phase
may be determined by the CP-violating nature of the effective potential
as well as the boundary condition in the broken phase region, which
is restricted by experiments at zero temperature.
To have the explicit form of $g(x)$,
we need to specify the model, to calculate the effective potential and
to solve the equations of motion with an appropriate boundary condition
which is consistent with experiments. This will be very important but
difficult process to estimate the baryon asymmetry quantitatively. We
do not execute this program here but assume two forms of $g(x)$ to
evaluate the chiral flux.
Note that it is $g^\prime (x)$, not $g(x)$ itself, that contributes
to $\Delta R$ as a factor of the integrand of it.
\footnote{\dag}{See Eqs.(2.25), (2.27) and (2.35) in I.}

In the first case, $g(x)$ is related to the real part $f(x)$ in a
simple manner:\par
\noindent (a)  $g(x)=\Delta\theta \, f^n(x)$ with $n=2$.
\footnote{\dag\dag}{$\Delta R=0$ for $n=1$ as pointed out in I.} \par
\noindent In the second case, $g^\prime (x)$ has the same thickness with that
of $f(x)$
and tends to zero
both in the symmetric and broken phase regions:\par
\noindent (b)  $g^\prime (x)=\Delta\theta \, {\rm sech} (x)$,
the CP-violating range being equal to the wall thickness.\par
\noindent These are free from any
 restriction from experiments, and give large $\mid \Delta R/\Delta \theta
\mid$. \par
%\noindent (a)  $g(x)=\Delta\theta \, f^n(x)$ with $n=2$.
%\footnote{\dag}{\dag}{$\Delta R=0$ for $n=1$ as pointed out in I.} \par
%ange being equal to the wall thickness. \par
%\noindent (a)  $g(x)=\Delta\theta \, f^n(x)$ with $n=2$.
%\footnote{\dag}{\dag}{$\Delta R=0$ for $n=1$ as pointed out in I.} \par

Fig.1 shows $\Delta R/\Delta\theta $ for the case (a).
For all forms of $g(x)$ and
$g'(x)$ we have analyzed, $\mid\Delta R\mid$ as a function of
$a/m_0$ decreases as $a/m_0$ decreases almost exponentially
 (thick walls). \par
Fig.2 illustrates $u$ dependence of $F^i_Q/u$
 for the case (a), which is found to be almost constant for $u \ltsim 0.5$.
\par
Fig.3 shows the contour plots of $F^i_Q/(u T^3)$ for $u\ltsim 0.5$
for the cases (a) and (b).
The both plots share a common pattern. Namely, they have the crest
lines along $m_0 \sim a$, and fall particularly rapidly for
$m_0 \gtsim a$
since $\Delta R$ itself decreases exponentially as
stated above.
The crest lines are sharply peaked at $m_0\sim a\sim T$. There we have
$$F^i_Q/(u T^3) \simeq 10^{-3}\,(Q^i_L-Q^i_R)\,\Delta\theta.
\eqno(3.3)
$$
\par
{\bf ---Figs.1,2,3---}
\chapter{Remarks on Electroweak Baryogenesis}

First we briefly review how the chiral charge flux transforms to the baryon
number
through the sphaleron process along the charge transport scenario${}^{3)}$.
We take the weak hypercharge $Y$ as the conserved charge in the symmetric
phase.
Fermion distributions near the bubble wall are changed by the hypercharge flux,
which are to be obtained by solving the Boltzmann equations.
Suppose that the initial states are in chemical equilibrium with $B=L=0$
through
the gauge and Yukawa interactions.
New chemical equilibrium emerges by injection of the hypercharge flux.
These situations are described by the chemical potentials $\mu$.
{}From the chemical equilibrium and initial conditions, the chemical potentials
for the baryon number ($\mu_B$) and the weak hypercharge ($\mu_Y$) are related
as
 $$\mu_B=-\mu_Y/6.\eqno(4.1)$$
The hypercharge density $ n_Y $ is expressed by the chemical potential
$ \mu_Y $ as
$$n_Y={T^2\over 6}{(1+2m)\over4}\mu_Y,\eqno(4.2)$$
where $ m $ is the number of Higgs doublets.
The transformed baryon density $\rho_B$ in the symmetric phase region is
obtained by integrating
the detailed balance equation,
$$\dot \rho_B=-{\Gamma_B\over T}\mu_B
           =-{\Gamma_B\over T}{6\over T^2}n_B(z,t),\eqno(4.3)$$
where $\Gamma_B$ is the sphaleron transition rate.
Considering that $n(z,t)$ depends only on the distance from the wall in
the rest frame of it,
we put $n(z,t)=n(z-ut)\equiv n(z')$.
%From the diffusion equations,$n^i_Q(z')$ is estimated as${}^{10)11)}$
%$$n^i_Q (z')\simeq F^i_Q exp(-{2\over\tau^i}z'),\eqno(4.4)$$
%where $\tau^i(=D^i/u)$ corresponds to the transport time of Ref.3) and
%$D^i$(quark)$\simeq 1/T$ and $D^i$(lepton)$\simeq (10^2\sim 10^3)/T$.
Then,
$$
\rho_B=-{\Gamma_B\over T}{6 \over T^2}\int_{-\infty}^{z/u}n_B(z-ut)dt
%$$={4\over{1+2m}}{\Gamma_B\over T^3}\int_{-\infty}^{z/u}n_Y(z-ut)dt$$
={4\over{1+2m}}{\Gamma_B\over T^3}{1\over
u}\int_0^{\infty}n_Y(z')dz'.\eqno(4.4)$$
The last integral is estimated as
$${1\over u}\int_0^{\infty}n_Y(z')dz'\simeq{F^i_Y\over u}\tau^i,\eqno(4.5)$$
%$$\simeq{4\over{1+2m}}{\Gamma_B\over T^3}{F^i_Y\over u}\int_0^{\infty}
%exp(-{2\over\tau^i}z')dz'$$
%$$\simeq{2\over{1+2m}}{\Gamma_B\over T^3}{F^i_Y\over u}{\tau^i}.\eqno(4.5)$$
where $\tau^i$ is the transport time within which the scattered fermions are
captured
by the wall${}^{3)}$.\par
If $\tau^i$ is given by $D^i/u$ with the diffusion constant $D^i$ ${}^{6)7)}$,
 it is estimated from
$D^i$(quark)$\simeq 1/T$ and $D^i$(lepton)$\simeq (10^2\sim 10^3)/T$. From
$\Gamma_B=\kappa\alpha^4_W T^4\simeq 10^{-6}T^4$ with $\kappa \simeq 0.1-1.0$,
we obtain
$${\rho_B \over s} \simeq 10^{-7}\times{F^i_Y\over u}\times{\tau^i\over T^2},
\eqno(4.6)$$
where the entropy density is given by
 $s=2\pi^2g_{\ast}T^3/45$ with $g_*\simeq 100$.

\par
As we saw in the previous section, $F_Y$ strongly depends on $a$ and $m_0$.
The effective potentials of the minimal standard model making use of the
high-temperature expansion suggest rather
thick bubble walls ($1/a =(20\sim40)/T$)${}^{8)9)10)}$.
We have for $T\simeq100$GeV and $u=0.1$,
$$
\eqalign{&\rho_B/s \simeq 10^{-18}\times\Delta\theta\quad \rm{(for\ top\
quark)},\cr
&\rho_B/s \simeq 10^{-13}\times\Delta\theta\quad \rm{(for\ \tau\ lepton)}.\cr
}
\eqno(4.7)$$
Some models, such as a two-Higgs-doublet model proposed by three of the authors
(K.F., A.K. and K.T.) without resorting to the high-temperature expansion
${}^{11)}$, predict
thin walls with $1/a\sim1/T$.
Then we expect rather large $F^i_Y$ which leads to
$$\rho_B/s=(10^{-9}\sim 10^{-10})\times\Delta\theta
\quad\rm{(for\ top\ quark\ and\ \tau\ lepton)}.\eqno(4.8)$$

Now some comments are in order.
\par
(1) $D^i$ could be estimated from the Boltzmann equations
and is expected to be $D^i=l^i/3$ where $l^i$ is the mean free path of the
fermion.
But there is a suggestion of an enhancement of
 $\tau$ by a factor of $O(10^3)$ due to the forward multiple
scattering
 effects${}^{3)}$.
If so, this leads to
$$\rho_B/s =(10^{-6}\sim 10^{-7})\times\Delta\theta,\eqno(4.9)$$
and would be able to explain the observed value.
\par
(2) As $u\rightarrow 0$, $\rho_B$ in (4.6) diverges because
$\tau^i(=D^i/u)\rightarrow\infty$. However, physical
considerations suggest that the limit $u\rightarrow 0$ leads to
a contradiction to the nonequilibrium condition.
Some discussions were made that there is a cutoff velocity
$u_{min} \sim 10^{-4} - 10^{-2}$ ${}^{7)9)}$ and
that $\rho_B\rightarrow 0$ as $u\rightarrow 0$ at $u < u_{min}$.
\par
(3) In the minimal standard model, the condition to suppress the
sphaleron transition
after the electroweak phase transition leads to the Higgs
mass bound $m_H \ltsim 50$ GeV
in contradiction to the experimental results.
Also in the model, the CP-violating effect emerges only in the KM matrix
and is too small to explain the observed value of $\rho_B/s$.
Extensions of the standard model by incorporating  two Higgs doublets
(including SUSY model)
 have the possibility to avoid the above upper bound of $m_H$.\par
In order to get $F^i_Y$ quantitatively, however, we must know
the spatial dependence of the CP-violating phase $\theta(x)$.
 We are trying  to solve nonlinear equations of motion
for $\theta(x)$  of the two-Higgs-doublet model ${}^{5)11)}$ under some
assumptions for the real part of the bubble wall.
\par
(4) A time-dependent  $\dot\theta(x,t)$ plays a role similar to
 the chemical potential and is called charge potential.
It influences the flux densities $f^s_{L(R)i}$ and
causes baryogenesis through the sphaleron process (spontaneous baryogenesis
scinario)${}^{3)}$.
The function $\Delta f^s_i=f^s_{Li}-f^s_{Ri}$
is  given from $\dot\theta(x,t)$ through the Boltzmann or diffusion equations.
For thick walls, the second terms in
 (2.2) and (2.7), which have been ignored in this paper, would
  become effective.
Some results under the classical and hydrodynamical
treatment were reported${}^{12)}$. \par
In conclusion,
we need to obtain the profile of the bubble wall and,
 in particular,
 the functional form of $\theta(x,t)$
 in order to estimate $F^i_Y$ quantitatively, which is the basic quantity
in any scinario of electroweak baryogenesis.

\par

\bigskip
{\bf References}
\item{1)} A.Sakharov, JETP Lett.{\bf 5}(1967),24.
\item{2)} V.Kuzmin, V.Rubakov and M.Shaposhnikov,
Phys.Lett.{\bf B155}(1985),36.
\item{3)} A.Nelson, D.Kaplan and A.Cohen,
Nucl.Phys.{\bf B373}(1992),453;\nextline A.Cohen, D.Kaplan and A.Nelson,
Ann.Rev.Nucl.Part.Sci{\bf 43}(1993),27.
\item{4)} K.Funakubo, A.Kakuto, S.Otsuki, K.Takenaga and F.Toyoda,
\nextline
Phys.Rev.{\bf D50}(1994),1105.
\item{5)} K.Funakubo, A.Kakuto, S.Otsuki, K.Takenaga and F.Toyoda,
\nextline
hep-ph/9407207.
\item{6)} M.Joyce, T.Prokopec and N.Turok, Phys.Lett.{\bf B338}(1994),269.
\item{7)} A.F.Heckler, Phys.Rev.{\bf D51}(1995),405.
\item{8)} M.Dine, R.Leigh, P.Huet, A.Linde and D.Linde, Phys.Rev.{\bf
D46}(1992),550.
\item{9)} B.-H.Liu, L.McLerran and N.Turok, Phys.Rev.{\bf D46}(1992),2668.
\item{10)} N.Turok and J.Zadrozny,Nucl.Phys.{\bf B369}(1992),729.
\item{11)} K.Funakubo, A.Kakuto and K.Takenaga, Prog.Theor.Phys.{\bf 91}
(1994),341.
\item{12)} M.Joyce, T.Prokopec and N.Turok, \nextline
preprint PUPT--94--1496(1994), hep-ph/9410282.

\bigskip

{\bf Figure Cations}
\par
\item{Fig.1.}$\Delta R/\Delta\theta$ in the case (a) {\it vs.} $p_L/m_0$ for
several $a/m_0$.
\item{Fig.2.} $F^i_Q/(uT^3(Q^i_L-Q^i_R)\Delta\theta)$ {\it vs.} $u$ at
$a/T=10, 1, 0.1$ in the case (a) with $a/m_0=5$.
\item{Fig.3.} Contour plot of $F^i_Q/(uT^3(Q^i_L-Q^i_R)\Delta\theta)$
at $u\ltsim 0.5$ as a function of the wall thickness $1/a$ and the
fermion mass $m_0$. (a) $g(x)=\Delta\theta \, f^2(x)$.
(b) $g'(x)=\Delta\theta \, $sech$x$.
\bye